\documentclass[aip,preprint,amsmath,amssymb,graphicx]{revtex4-1}

\pdfoutput=1

\draft 

\usepackage{siunitx}

\usepackage{graphicx}

\usepackage{mathrsfs}

\usepackage[mathlines]{lineno}

\usepackage[utf8]{inputenc}
\usepackage[T1]{fontenc}
\usepackage{mathptmx}
\usepackage{etoolbox}

\begin{document}

\title{Thin film notch filters as platforms for biological image processing}

\author{Shaban B. Sulejman}
\email{sulejmans@unimelb.edu.au}
\affiliation{ARC Centre of Excellence for Transformative Meta-Optical Systems, School of Physics, The University of Melbourne, Victoria 3010, Australia.}

\author{Niken Priscilla}
\email{niken.priscilla@unimelb.edu.au}
\affiliation{ARC Centre of Excellence for Transformative Meta-Optical Systems, School of Physics, The University of Melbourne, Victoria 3010, Australia.}

\author{Lukas Wesemann}
\affiliation{ARC Centre of Excellence for Transformative Meta-Optical Systems, School of Physics, The University of Melbourne, Victoria 3010, Australia.}

\author{Wendy S. L. Lee}
\affiliation{ARC Centre of Excellence for Transformative Meta-Optical Systems, School of Physics, The University of Melbourne, Victoria 3010, Australia.}
\affiliation{ARC Centre of Excellence for Transformative Meta-Optical Systems, Department of Electrical and Electronic Engineering, The University of Melbourne, Victoria 3010, Australia.}

\author{Jieqiong Lou}
\affiliation{School of Physics, The University of Melbourne, Victoria 3010, Australia.}

\author{Elizabeth Hinde}
\affiliation{School of Physics, The University of Melbourne, Victoria 3010, Australia.}

\author{Timothy J. Davis}
\affiliation{ARC Centre of Excellence for Transformative Meta-Optical Systems, School of Physics, The University of Melbourne, Victoria 3010, Australia.}

\author{Ann Roberts}
\affiliation{ARC Centre of Excellence for Transformative Meta-Optical Systems, School of Physics, The University of Melbourne, Victoria 3010, Australia.}

\begin{abstract}

Many image processing operations involve the modification of the spatial frequency content of images. Here we demonstrate object-plane spatial frequency filtering utilizing the angular sensitivity of a commercial spectral bandstop filter. This approach to all-optical image processing is shown to generate real-time pseudo-3D images of transparent biological and other samples, such as human cervical cancer cells. This work demonstrates the potential of non-local, non-interferometric approaches to image processing for uses in label-free biological cell imaging and dynamical monitoring. 

\end{abstract}

\maketitle



\section{Introduction}

Transparent objects, including most biological cells, interact weakly with light resulting in little contrast in conventional bright field microscopy. However, spatial variations in their morphology and optical properties introduce local spatial phase variations onto light transmitted through them. In the simplest case, this can be characterized by a transmission function \(O(x,y) \approx O_0 e^{i\varphi(x,y)}\). While the amplitude \(O_0\) is approximately spatially invariant, the shape and refractive index information is contained in the phase function \(\varphi(x,y)\). Such phase variations cannot be directly sensed by conventional cameras and requires indirect detection. Popular optical phase visualization methods include Zernike \cite{Zernike1942} and differential interference contrast microscopy \cite{Lang1982}. However, these can require expensive components or Fourier plane access that increases system complexity and size. Digital methods include ptychography \cite{Balaur2021, Horstmeyer2016, Giewekemeyer2010}, the use of the transport of intensity equation \cite{Engay2021,Bostan2016,Li2016} or phase retrieval algorithms such as the Gerchberg-Saxton \cite{Gerchberg1972} and Fienup algorithms \cite{Fienup1978}. However, these are limited by their extensive computational requirements. 

All-optical, object-plane image processing offers a non-interferometric compact alternative for phase visualization. It is enabled by 2D space-invariant linear optical systems, such as thin films \cite{Wesemann2019, Zhu2017}, with angular responsivities that directly filter the spatial frequency of wavefields \cite{Davis2021}. Unlike common computational or all-optical methods utilizing the classical \(4f\)-configuration \cite{Goodman1996}, it avoids optical phase information losses, energy consuming post processing and bulky configurations associated with accessing Fourier planes. The importance of compact optical systems for all-optical, object-plane image processing is motivated by the potential for integration into portable devices. This can have applications as diverse as mobile diagnostics, environmental monitoring and remote sensing.

To explain how a device exhibiting angular dispersion can perform image processing, we ignore any polarization effects for simplicity. In this case, the impact of object-plane Fourier filtering on the spatial frequency spectrum of the field can be described by an optical transfer function \(\mathcal{H}(k_x, k_y)\) \cite{Davis2019}. By taking the \(z\)-axis as the optical axis, \(k_{x}\) and \(k_{y}\) denote the transverse spatial frequency components of the wave-vector \(\Vec{k} = (k_x, k_y, k_z)\) and \(k_z = \sqrt{|\Vec{k}|^2 - k_x^2 - k_y^2}\). The transfer function relates the processed output to the input by the convolution theorem,
\begin{equation}
\label{eq:otf-conv-thm}
    E_{ \text{out} }(x,y,z) = \mathscr{F}^{-1} \left\{ \mathcal{H}(k_x, k_y) \Tilde{E}_{ \text{in} }(k_x, k_y; z) \right\} (x, y) \ ,
\end{equation}
where \(\mathscr{F}\) denotes the Fourier transform, \(E\) represents any component of the electric field and \( \Tilde{E}_{\text{in}} \equiv \mathscr{F}\left\{ E_{ \text{in} } \right\}\). For example, high-pass filters block low spatial frequencies to eliminate unscattered field components for edge detection \cite{Roberts2018}, which is fundamental to data compression \cite{Guo2019} and machine vision \cite{Brosnan2004, Park2021}. A notable sub-class is that of linear optical transfer functions, i.e. \(\mathcal{H} \propto k_x\) or \(\mathcal{H} \propto k_y\). These can directly translate phase gradients into intensity variations through spatial differentiation in the case of transparent objects to permit phase visualization \cite{Wesemann2019}. The influence of polarization can be incorporated into this approach by utilizing a \(2 \times 2\) transfer function dyadic tensor. 

The angular sensitivity provides the mechanism for image processing through the correspondence between angles of incidence and spatial frequencies. This is given by representing the spatial frequency components in spherical coordinates \cite{Davis2021},
\begin{equation}
    \label{eq:spatial-freq-sph-coords}
    \begin{gathered}
        k_x = k_0 \sin{\theta} \cos{\phi} \\
        k_y = k_0 \sin{\theta} \sin{\phi} \\
        k_z = k_0 \cos{\theta} \ ,
    \end{gathered}
\end{equation}
where \(k_0=|\Vec{k}|\) is the wavenumber, while  \((\theta, \phi)\) are the polar and azimuthal propagation angles of plane waves with respect to the \(z\)-axis. Given that optical transfer functions represent plane wave responses in \(k\)-space, and that light can be decomposed into weighted plane waves by the spatial Fourier transform \cite{Davis2021,Goodman2000}, then devices exhibiting angular dispersive transmission are capable of object-plane image processing. 

Recently, meta-optical devices have attracted considerable attention as ultra-compact image processors \cite{Silva2014}. For example, Zhou et al. \cite{Zheng2020} employed photonic crystals for edge detection of organic samples, while Wesemann et al. \cite{Wesemann2021} obtained phase contrast images of human cancer cells using a resonant waveguide grating. Other approaches have involved Mie \cite{Wan2020,Komar2021} or Fano \cite{Cordaro2019} resonances, photonic spin-orbit coupling effects \cite{Zhu2018,Wesemann2022} and bound states in the continuum \cite{Pan2021}. With this rapidly growing interest in meta-optics, it is timely to consider other optical elements capable of performing an equivalent role. Earlier works have investigated various structures for edge detection such as volume hologram filters \cite{Case1979}, Fabry-Pérot etalons \cite{Indebetouw1980}, detuned interference filters \cite{Molesini1981, Molesini1982, Cetica1982}, acousto-optic modulators \cite{Indebetouw1984} and gratings \cite{Angell1985, Peri1985, Marquez2003}. With the exception of Fourier plane phase contrast methods, the emergence of successful digital methods for image processing held back further progress in all-optical techniques.

Here we demonstrate the use of a commercially available thin film spectral notch filter applied to phase contrast imaging of weakly absorbing phase objects. Notch filters are band-stop filters commonly employed in various types of spectroscopy to remove temporal frequencies over a specific range \cite{Marques2012, Antonacci2022}. We show that the angular dispersion of the filter's rejection band produces a high-pass spatial frequency filter at the operating wavelength. Introducing a phase bias by tilting the filter with respect to the optical axis can produce pseudo-three dimensional images similar to those obtained in differential interference contrast microscopy. Moreover, the contrast generated is determined by the rotation axis and angle. We demonstrate enhanced contrast imaging of wavefields introduced by a spatial light modulator and unstained biological samples, including human cervical cancer cells. Our results confirm real-time phase contrast imaging without post-processing, permitting direct imaging with either a camera or the eye. This method provides a novel all-optical approach for biological and other image processing using off-the-shelf band-stop devices. It has the potential for developments in machine vision, biological imaging and dynamical monitoring.


\section{Experimental Methods \& Results}

\subsection{Device Performance}

The device investigated here is a commercially obtained notch filter (Thorlabs NF633-25), which has a specified central operating wavelength of \SI{633}{\nano\meter} and band-width of \SI{25}{\nano\meter} at normal incidence. Transmission spectra as a function of angle of incidence were experimentally measured using the configuration in Fig. \ref{fig:Spectra_OTF_config}, with details provided in the supplementary information (\S S2A). The results (Fig. \ref{fig:Spectra_OTF_results}(a)) obtained using circularly polarized light are consistent with the manufacturer's specifications. Meanwhile, the results for \(p\)- and \(s\)-polarized light are given in the supplementary material (\S 2A). Noting the correspondence between \((\theta, \phi)\) and \((k_x, k_y)\), it is found that the notch filter suppresses low spatial frequencies associated with near-normal incidence angles at the band-stop wavelength. 

\begin{figure}[htb]
   \centering
    \includegraphics[width=\linewidth]{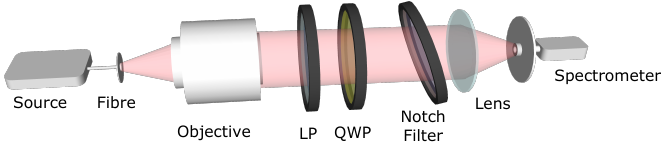}
    \caption{\label{fig:Spectra_OTF_config}The experimental configuration used to capture transmission spectra of the notch filter. Here, LP and QWP denote linear polarizer and quarter wave-plate, respectively.}
\end{figure}

Modulation transfer functions \(|\mathcal{H}(k_x, k_y)|\) were mapped from the measured transmission spectra for \(p\)-, \(s\)- and circular polarizations. Line profiles along \(k_y=0\) (Fig. \ref{fig:Spectra_OTF_results}(b)) at the band-stop wavelength exhibit approximately polarization-insensitive, high-pass behaviour with a suppression zone of numerical aperture (NA) range \(\sim 0.2\). This is produced by the blue-shifting of the band-stop wavelength with increasing angle of incidence. 

\begin{figure}[htb]
   \centering
    \includegraphics[width=\linewidth]{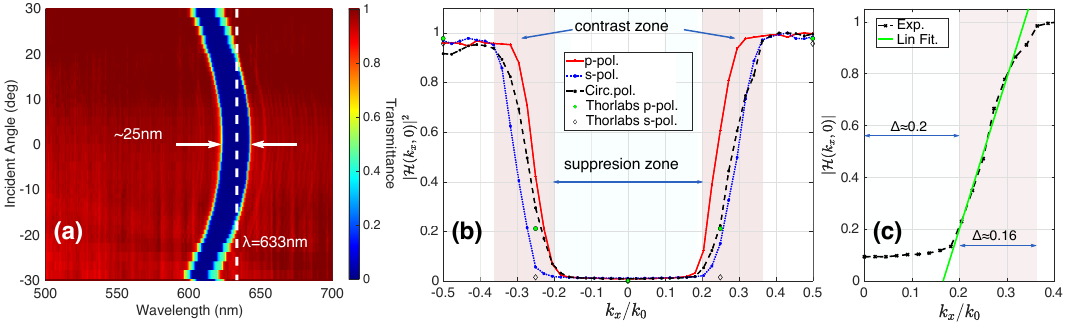}
    \caption{\label{fig:Spectra_OTF_results}The experimental transmission spectra obtained by incrementally rotating the filter is given in (a). The modulation transfer function along \(k_y=0\) produces the plot shown in (b) for various polarizations, which is compared to data provided by the manufacturer. Linear fitting of the modulation transfer function in the contrast zone is given in (c).}
\end{figure}

A significant feature of the device is the region of approximate linear dependence on \(k_x\) from zero to near-unity transmission beyond the suppression zone. This is supported by linear fitting (Fig. \ref{fig:Spectra_OTF_results}(c)) from \(k_x/k_0 \sim 0.2\) to \(0.3\), represented by the relation \(A k_x/k_0 + B\). The polyfit function and curve fitting toolbox in MATLAB were used to obtain the fitting parameters and their error intervals as \(A = 5.9 \pm 0.83\) and \(B = -0.97 \pm 0.22\). Rotating the filter by \SI{12}{\degree} about a line perpendicular to the optical axis shifts operation to \(k_x/k_0 \approx \pm 0.2\) to access this region, referred to here as the contrast zone. Such a linear optical transfer function substituted into Eq. \eqref{eq:otf-conv-thm} approximately produces the spatial derivative around an intensity offset along the axis of rotation. This can generate intensity images directly related to phase gradients \cite{Davis2021}, given by
\begin{equation}
    \label{eq:phase-gradient}
    \nabla O(x,y) \approx i O_0 \left( \nabla \varphi(x,y) \right) e^{i\phi(x,y)} \ .
\end{equation}
Consequently, intensity contrast is created in regions where the phase is varying along the relevant direction by the direct relationship in Eq. \eqref{eq:phase-gradient}. Operating near the edge of the contrast zone removes unscattered field components leaving only relatively large phase gradients that enhance edges. However, operating within the contrast zone at a rotation angle of \SI{14}{\degree}, corresponding to \(k_x/k_0 \approx 0.24\), preserves some unscattered components and enables discrimination between positive and negative phase gradients. These manifest as different grayscale levels above or below the shifted \(k\)-space origin, respectively. Therefore, operation within the contrast zone enables phase visualization by converting phase gradients to intensity. This includes information about the sign of the gradient to produce pseudo-3D images. Phase gradients along different directions can also be visualized by changing the axis of rotation. These form the key underpinnings of this paper to describe the capacity of notch filters for phase visualization. 


\subsection{Phase Contrast Imaging}

All-optical image processing was experimentally performed on various samples with the notch filter using circularly polarized, \SI{635}{\nano\meter} laser light. Edge detection of amplitude objects was first demonstrated on a USAF resolution test target, with details and results provided in the supplementary information (\S S2B). Phase imaging was then performed using the configuration (Fig. \ref{fig:SLM_imaging_config}) detailed in the supplementary information (\S S2C). A computer-controlled spatial light modulator emulated phase profiles of human red blood cells (Fig. \ref{fig:SLM_imaging_results}(a)), which were modelled using the properties obtained from Ref. \citenum{Evans1972}. Collimated light reflected from the spatial light modulator passed through the notch filter, which was rotated within the focal plane between paired microscope objectives. A camera then captured the transmitted images.

\begin{figure}[htb]
   \centering
    \includegraphics[width=\linewidth]{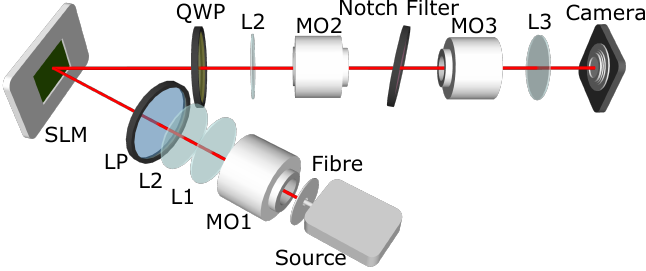}
    \caption{\label{fig:SLM_imaging_config}The configuration for image processing with a spatial light modulator (SLM), where L and MO denote lenses and microscope objectives, respectively.}
\end{figure}

The results are presented in Fig. \ref{fig:SLM_imaging_results}(b)-(e), where simulated and experimental images are compared. An experimental control image without the filter (Fig. \ref{fig:SLM_imaging_results}(b)) displays poor contrast, rendering the object virtually invisible as expected for weakly absorbing phase objects. Simulated and experimental phase images (Fig. \ref{fig:SLM_imaging_results}(c) and (d)) are obtained by introducing the filter at a rotation angle of \SI{14}{\degree}. These preserved some unscattered contributions leading to the appearance of pseudo-3D phase contrast. Moreover, the simulated phase image (Fig. \ref{fig:SLM_imaging_results}(c)) is consistent with experiment (Fig. \ref{fig:SLM_imaging_results}(d)). In both cases, regions where a change in phase is present, i.e. \(\nabla \varphi(x,y) \neq 0\), are transformed to intensity variations arising from otherwise invisible phase modulations. Moreover, ringing artifacts associated with the Gibbs phenomenon can also be seen. The resultant images additionally possess the capacity to distinguish between positive and negative phase gradients. Furthermore, line profiles (Fig. \ref{fig:SLM_imaging_results}(e)) exhibit intensity variations associated with phase variation introduced into the field by the sample. Some intensity artifacts can also be seen, such as those arising from amplitude variations in the field.

\begin{figure}[htb]
   \centering
    \includegraphics[width=0.6\linewidth]{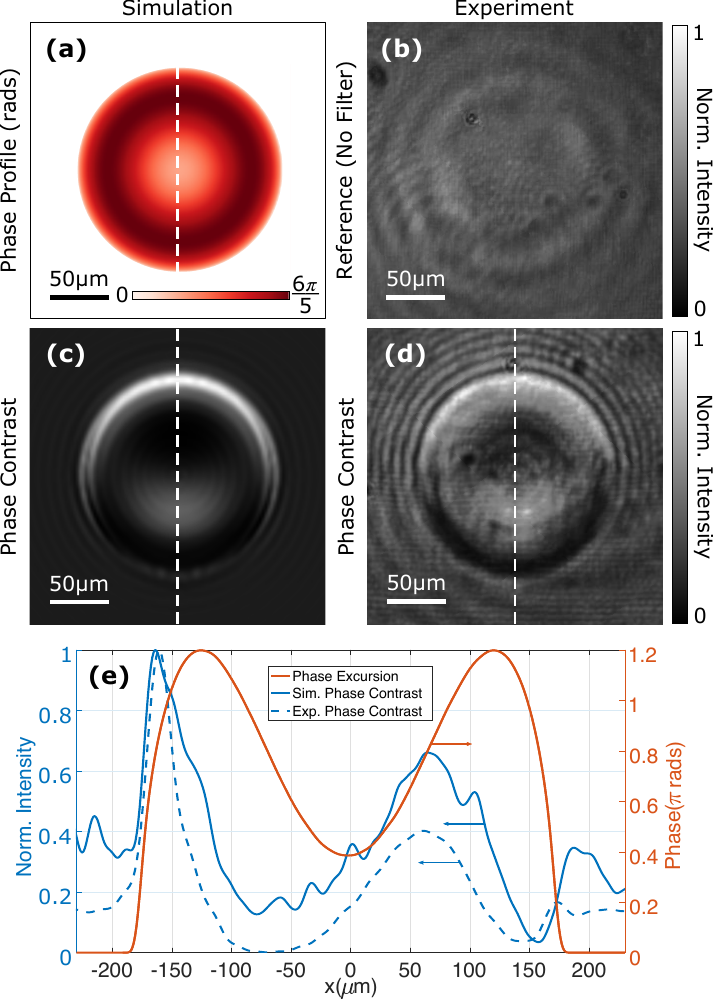}
    \caption{\label{fig:SLM_imaging_results} A red blood cell (a) was emulated by a spatial light modulator with a phase excursion of \(6\pi/5\). Simulated and experimental images normalized to their brightest pixels are given in (b)-(d), comprising control (b), simulated (c) and experimental (d) phase contrast images. These are supported by line profiles in (e) along the dashed lines shown in (b)-(d).}
\end{figure}

\begin{figure}[htb]
   \centering
    \includegraphics[width=0.8\linewidth]{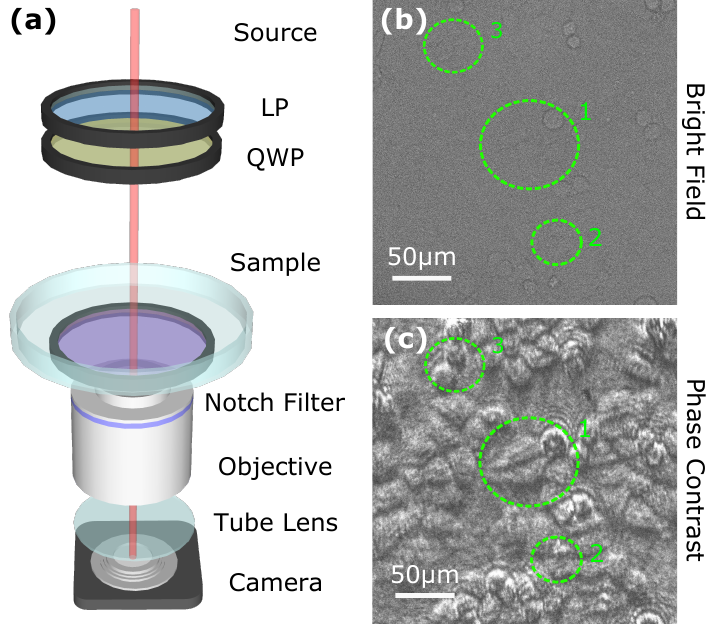}
    \caption{\label{fig:Bio-phase_imaging} An experimental schematic for biological phase imaging is given in (a) and outlined in the supplementary information (\S S2D). A bright field image of the HeLa cells obtained without the filter is given in (b). The corresponding phase contrast image obtained using the filter is given in (c).}
\end{figure}

To illustrate potential microscopy applications, phase contrast microscopy images were obtained of biological samples with weak amplitude contrast. This was performed using an inverted microscope configuration depicted in Fig. \ref{fig:Bio-phase_imaging}(a). Moreover, human cervical cancer (HeLa) cells were used as the sample and the preparation steps are outlined in the supplementary information (\S S2E). Their bright field image (Fig. \ref{fig:Bio-phase_imaging}(b)) displays little to no contrast. However, placing the notch filter immediately beneath the sample enabled visualization of the phase variations when illuminated with \SI{635}{\nano\meter} laser light. The phase contrast image (Fig. \ref{fig:Bio-phase_imaging}(c)) obtained within the contrast zone contains morphological detail absent in the corresponding bright field image, such as those highlighted within the green curves. The contrast produced exhibits significantly enhanced phase visualization compared to the relatively low contrast bright field image. Cellular thickness and local refractive index deviations are accented to discriminate the cells from their background. A conventional differential interference contrast image was also taken of the same regions, which is given in the supplementary information (\S S2D) as a baseline for comparison.

\section{Discussion}

Inspired by the recent surge in interest towards meta-optical imaging, these results represent the first demonstration of biological image processing using a commercially available filter. The spatial frequency content of images was directly altered by the notch filter in the object plane without needing to access the Fourier plane. The transfer function exhibited the required behaviour for high-pass filtering enabling edge detection, in addition to approximately linear regions allowing phase visualization through spatial differentiation. Rotating the filter enabled access to these regimes to offset the Fourier origin in \(k\)-space. Phase gradients were converted into intensity variations to generate images with significantly enhanced contrast. As a result, notch filters offer a new approach to non-interferometric phase visualization through image processing without bulky or expensive components. 

Commercial notch filters offer a relatively cost efficient and readily available option for phase visualization compared to alternative optics. Although the band-stop region was limited within the visible band, various filters have been designed with coverage across the electromagnetic spectrum. For example, Yuan et al. \cite{Yuan2020} presented a photonic filter employable at the absorption band of acetylene gas. Others include terahertz \cite{Bark2021}, infrared \cite{Kwon2004}, microwave \cite{Long2017, Gertler2022} and commercial ultra-violet filters. These hold potential for phase imaging across the spectrum using the methods outlined in this paper, with applications in machine vision and biological imaging beyond the visible range. Live biological monitoring is possible with the notch filter dynamically capturing phase information of specimens. Further applications include integration into conventional CMOS sensors to form spatial frequency and phase-sensitive detectors. Incorporating notch filters into their detection planes would introduce spatial frequency selectivity and phase variations in otherwise indiscriminate, intensity-based devices.

Although phase contrast was produced by the notch filter, its behaviour in the contrast zone was not perfectly linear. Furthermore, its restricted numerical aperture limits the range of samples to which it could be applied. The large angular range of the suppression zone also restricts contrast to sufficiently sharp features. Not only is the range of objects limited, but it necessitates rotation of the notch filter. Finally, our theoretical and numerical modelling neglected the non-negligible notch filter thickness, which can introduce perturbations into the beam path. This can lead to the appearance of aberrations such as those in the edge enhanced images in the supplementary information (\S S2B). However, the success of experiments performed here provide confidence that custom thin film devices could be developed with reduced thicknesses to minimize aberrations and tailor the transfer function.


\section{Conclusion}

In conclusion, this paper has demonstrated real-time, all-optical, object-plane image processing using a commercial spectral notch filter. Spectroscopic measurements verified angle dispersive band-rejection necessary for high-pass spatial frequency filtering and phase contrast imaging. Edge detection was realizable through the suppression zone where unscattered field components were removed. Meanwhile, offsetting to approximately linear regions within the contrast zone produced phase contrast images. Unstained biological samples, including human cervical cancer cells, imposing otherwise invisible phase modulations were visualized by the filter. Given its real-time capabilities, this has significant implications in label-free biomedical imaging \cite{Park2018} including medical diagnostics, non-invasive micro-organism growth and dynamical monitoring. The results open developmental possibilities in extending to beyond the visible band and constructing monolithic spatial frequency-sensitive cameras for commercialization.


\section*{Funding}

This research was funded by the Australian Government through the Australian Research Council Centre of Excellence grant (CE200100010). S.B.S also gratefully acknowledges the support of the Ernst \& Grace Matthaei Scholarship and the Australian Government Research Training Program Scholarship.

\section*{Supplementary Material}

Supporting information is available in the supplementary material, including the underlying theory and detailed experimental methods.

\bibliography{References.bib}

\newpage


\setcounter{section}{0}
\renewcommand{\thesection}{S\arabic{section}}


\newpage

\centering
\textbf{\large Supplementary Material}



\section{Theoretical Modelling}

\subsection{Phase Visualization via Spatial Frequency Filtering}

The core idea underlying this work is to directly modify spatial frequencies of wavefields without accessing Fourier planes to enable phase extraction. The essential feature of suitable devices is a sensitivity to angles of incidence. As explained in the main article, this associates with a selective transmission to spatial frequencies necessary for image processing. 

Image processing for phase visualization can be explained using classical scalar electromagnetic theory \cite{Roberts2018SI}. Polarization effects can be accounted through tensorial analysis \cite{Davis2021SI}, but are ignored here for simplicity. Suppose that the source electric field can be modelled as monochromatic plane waves \(E_s(x,y,z)\) of wavelength \(\lambda\) and wavevector \(\Vec{k} = (k_x, k_y, k_z)\). With the \(z\)-axis as the optical axis, \(k_{x}\) and \(k_{y}\) denote the transverse spatial frequency components and \(k_0 \equiv |\Vec{k}| = \frac{2\pi}{\lambda}\) is the wavenumber. The source illuminates a weakly absorbing object located in the \(z=0\) plane, which is assumed as optically thin in the Kirchhoff approximation. Its transmitted light can be modelled by
\begin{equation}
\label{eq:obj-field}
    E_{ \text{obj} }(x,y,0) = O(x,y) E_s(x,y,0) \ ,
\end{equation}
where the object transmission function is approximately given by
\begin{equation}
\label{eq:obj-trans-func}
    O(x,y) \approx O_0 e^{i \varphi(x,y)} \ .
\end{equation}
Its intensity image \(|O(x,y)|^2 = |O_0|^2\) has little to no contrast by construction and is therefore referred to as a phase object. Therefore, its visualization requires access to the phase function \(\varphi(x,y)\) that contains its shape information. This is made possible through optical systems, such as thin films \cite{Zhu2017SI, Wesemann2019SI}, described by specific optical transfer functions that can extract \(\varphi(x,y)\) from the complex exponential of Eq. \eqref{eq:obj-trans-func}. To shed light on this, we note the following property for the Fourier transform of derivatives,
\begin{equation}
    \label{eq:FT-derivatives}
    \mathscr{F} \left\{ \frac{\partial^m E}{\partial x^m} + \frac{\partial^m E}{\partial y^m} \right\} (k_x, k_y) = i^m ( k_x^m + k_y^m ) \mathscr{F}\{E(x, y)\}(k_x, k_y) \ ,
\end{equation}
where \(\mathscr{F}\) denotes the Fourier transform and \(m \in \mathbb{N}\). As in Eq. (1) of the main article, the object light filtered by the notch filter is related to its optical transfer function \(\mathcal{H}(k_x, k_y)\),
\begin{equation}
\label{eq:img-proc-otf}
    E_{ \text{out} }(x,y,0) = \mathscr{F}^{-1} \left\{ \mathcal{H}(k_x, k_y) \mathscr{F}\left\{ E_{ \text{obj} }(x',y',0) \right\}(k_x,k_y) \right\} (x, y) \ ,
\end{equation}
where the notch filter is taken to immediately succeed the object. By comparing Eqs. \eqref{eq:FT-derivatives} and \eqref{eq:img-proc-otf}, the \(m^{th}\)-order derivatives of the input field are produced if the optical transfer function is given by
\begin{equation}
    \label{eq:spat-diff-otf}
    \mathcal{H}(k_x, k_y) = \begin{cases}
    i^m k_{x}^m, \ &\text{for x-differentiation}  \\
    i^m k_{y}^m, \ &\text{for y-differentiation}
    \end{cases} \ ,
\end{equation}
up to a multiplicative constant. It was shown in the main article that the transfer function exhibited approximately linear variation within the contrast zone. The notch filter is therefore capable of approximate first-order spatial differentiation in this region required for phase contrast imaging. Substituting a linear optical transfer function into Eq. \eqref{eq:img-proc-otf} and taking \(E_s(x,y,0) = \text{const.}\) produces Eq. (3) of the main article, given by 
\begin{equation}
    \label{eq:phase-gradient_SI}
    \nabla O(x,y) \approx i O_0 \left( \nabla \varphi(x,y) \right) e^{i\phi(x,y)} \ .
\end{equation}
Phase gradients of the object are translated into measurable intensity variations described by the amplitude term in Eq. \eqref{eq:phase-gradient_SI}. Thus, its intensity image is now given by 
\begin{equation}
    \label{eq:phase-contrast-image}
    |\nabla O(x,y)|^2 \approx |\nabla \varphi(x,y)|^2 \ ,
\end{equation}
with contrast generated where \(\nabla \varphi \neq 0\). This holds for approximately pure phase objects, with the gradient taken along the axis of rotation. However, non-negligible amplitude variations in \(O(x,y)\) would produce additional amplitude derivatives in Eq. \eqref{eq:phase-gradient_SI}. Although, it is possible to subtract these by using measured intensities in the absence of the filter to isolate the phase contribution.

\subsection{Simulations}

Numerical simulations were performed in Python by implementing the theoretical model presented in \S S2.1. Our experimentally measured modulation transfer function and the phase response provided the manufacturer were combined to model the optical response of the notch filter.


\section{Experimental Methods}


\subsection{Transfer Function Determination}

Spectroscopic measurements of the notch filter (Thorlabs NF633-25) were made using the experimental configuration depicted in Fig. 1 of the main article. The filter was \SI{3.5}{\milli\meter} thick and comprised tantalum pentoxide (Ta\(_2\)O\(_5\)) and silicon dioxide (SiO\(_2\)) thin films on a fused quartz substrate. White light from a fibre-coupled (Thorlabs SM600 single-mode fibre) halogen lamp (Ocean Insight HL-2000-FHSA) was collimated by a microscope objective (Nikon UPlanFL 20x/0.15NA). This light was polarized by a linear polarizer (Thorlabs LPVIS050-MP) to produce \(p\) or \(s\)-polarized light. Circularly polarized light was produced by introducing an additional quarter waveplate (Thorlabs AQWP05M-600) before illuminating the notch filter. A plano-convex lens (Thorlabs LA1068-A, \(f=\SI{75}{\milli\meter}\)) focused the transmitted light onto a port that was fibre-coupled (Thorlabs M15L01) to a spectrometer (Ocean Insight QE6500). Measurements of transmission spectra were made for incident angles ranging from \SI{-30}{\degree} to \SI{+30}{\degree} by incrementally rotating the filter by \SI{2}{\degree}. 

\begin{figure*}[htb]
   \centering
    \includegraphics[width=0.8\linewidth]{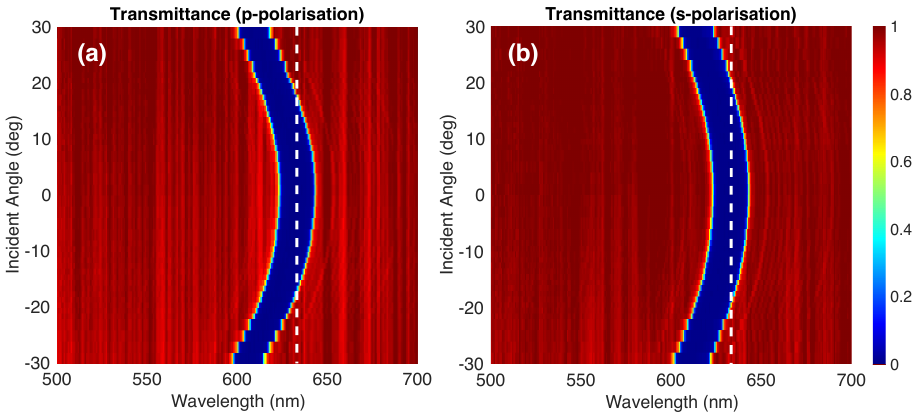}
    \caption{\label{fig:Spectra}Transmission spectra of the notch filter obtained with \(p\)- and \(s\)-polarized light are given in (a) and (b), respectively. These exhibit similar behaviour with suppression around \SI{633}{\nano\meter} (dashed lines) that blue-shifts with increasing incident angle.}
\end{figure*}

Spectra were collected for \(p\)-, \(s\)- and circularly polarized light. The former two are given in Fig. \ref{fig:Spectra}, while the latter is given in Fig. 2(a) of the main article. Each of these confirm suppression about the band-stop wavelength and its angular dispersion.


\subsection{Edge Detection}

\begin{figure}[htb]
   \centering
    \includegraphics[width=0.85\linewidth]{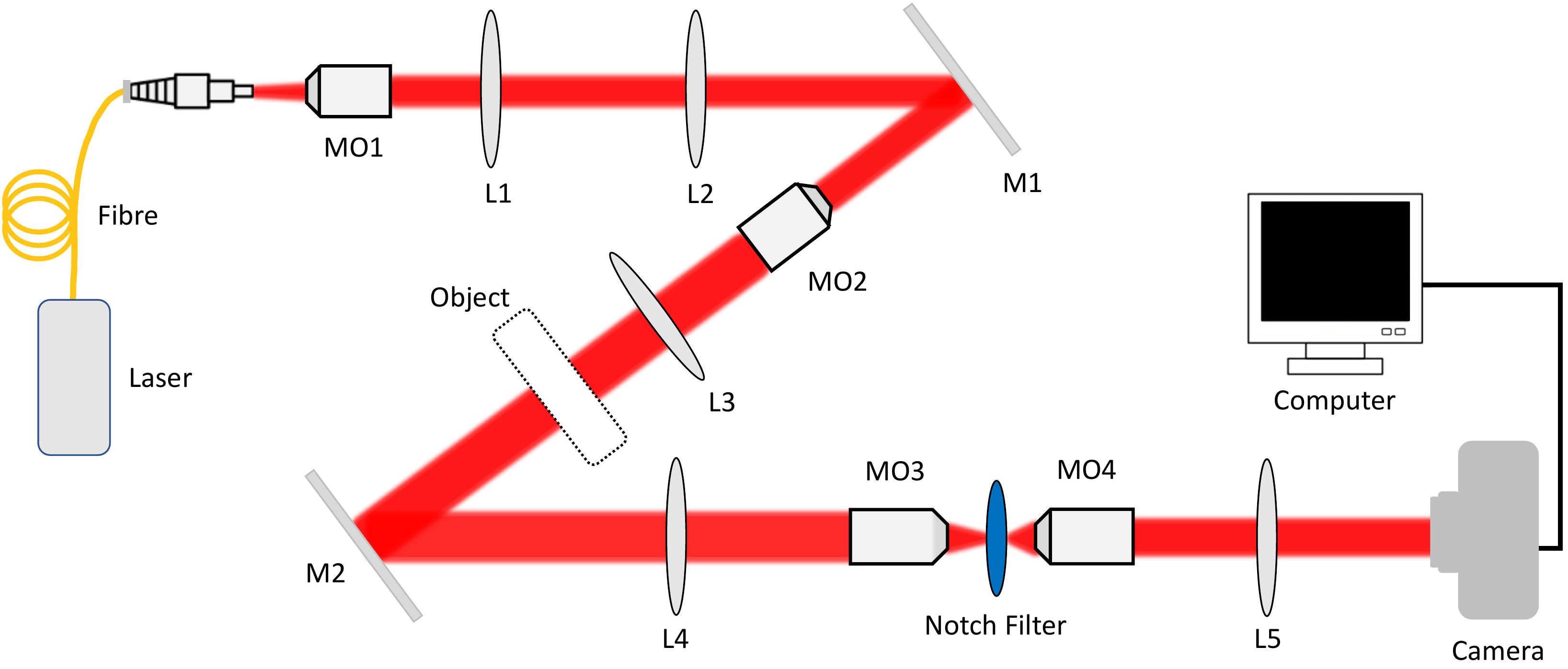}
    \caption{\label{fig:Edge-detect-config}A schematic of the experimental configuration for edge detection with a notch filter is shown. Here L, MO and M each denote lenses, microscope objectives and mirrors, respectively.}
\end{figure}

All-optical amplitude image processing was experimentally performed to demonstrate edge detection with the notch filter. As depicted in Fig. \ref{fig:Edge-detect-config}, fibre-coupled (Thorlabs SM600) laser light (Thorlabs S1FC635 \SI{635}{\nano\meter}) was collimated by a microscope objective (Nikon LU Plan 5x/0.15NA) and two lenses (Thorlabs LB1901-B \(f=\SI{75}{\milli\meter}\) \& Thorlabs LB1761-B \(f=\SI{25.4}{\milli\meter}\)). Light reflected from a mirror was projected onto the the largest regions of a USAF resolution test target (Thorlabs R2L2S1N1 NBS 1963A) using a second microscope objective (Olympus UPlanFI 20x/0.50NA) and lens (Thorlabs LA1509-A \(f=\SI{100}{\milli\meter}\)). Upon reflection from a second mirror, a fourth lens (Thorlabs LA1433-A \(f=\SI{150}{\milli\meter}\)) allowed image filtering with the notch filter between paired microscope objectives (Nikon UPlanFI 20x/0.4NA \& Nikon LU Plan 50x/0.55NA). Finally, a fifth lens (Thorlabs LA1131-A \(f=\SI{50}{\milli\meter}\)) projected the filtered light onto a camera (Thorlabs DCC1645C).

\begin{figure*}[htb]
   \centering
    \includegraphics[width=\linewidth]{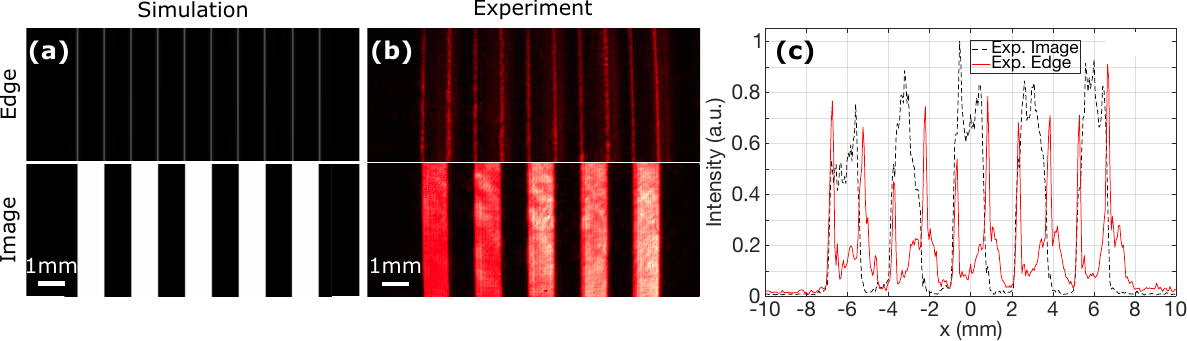}
    \caption{\label{fig:Edge-results}Edge-enhanced images of USAF test target bars overlayed on their unfiltered images are given in (a) and (b). Line profiles in (c) depict suppression within the uniform regions of the bars.}
\end{figure*}

Given in Figs. \ref{fig:Edge-results}(a) and (b), the simulated and experimental results demonstrate edge enhancement produced by the notch filter. The uniform intensity regions associated with unscattered field components were removed while retaining only the edges. This is supported by line profiles given in Fig. \ref{fig:Edge-results}(c). Some slight asymmetries are present in the experimental edge-enhanced image owing to small rotations of the filter.

\subsection{Phase Imaging}

All-optical image processing for phase visualization was performed using the configuration given in Fig. 3 of the main article. Fibre-coupled (Thorlabs SM600) laser light (Thorlabs S1FC635, \SI{635}{\nano\meter}) was collimated by a microscope objective (Nikon LU Plan 5x/0.15NA) and two plano-convex lenses (Thorlabs LA1027-A \(f=\SI{75}{\milli\meter}\) and Thorlabs LA1509-A \(f=\SI{100}{\milli\meter}\)). Required for its operation, a linear polarizer (Thorlabs LPVIS050-MP) ensured that linearly polarized light illuminated a reflective spatial light modulator (Holoeye Pluto-2-VIS-014 LCOS-SLM). Comprising 1920 \(\times\) 1080 liquid crystal pixels of \SI{8}{\micro\meter} pitch, it prescribed computer-generated phase profiles of human red blood cells. The reflected light from the SLM passed through a quarter wave-plate (Thorlabs AQWP05M-600) to circularly polarize it. A third plano-convex lens (Thorlabs LA1433-A \(f=\SI{150}{\milli\meter}\)) and microscope objective (Olympus PlanN 20x/0.4NA) together formed a telescope that de-magnified the SLM image onto the notch filter in its focal plane. To access the contrast zone, the latter was longitudinally rotated by \SI{14}{\degree} using a rotation mount. A second, identical microscope objective (Olympus PlanN 20x/0.4NA) and a fourth plano-convex lens (Thorlabs LA1131-A \(f=\SI{50}{\milli\meter}\)) re-magnified and focused the filtered image onto a camera (Thorlabs DCC1645C). 


\subsection{Phase Microscopy}

\begin{figure*}[htb]
   \centering
    \includegraphics[width=0.9\linewidth]{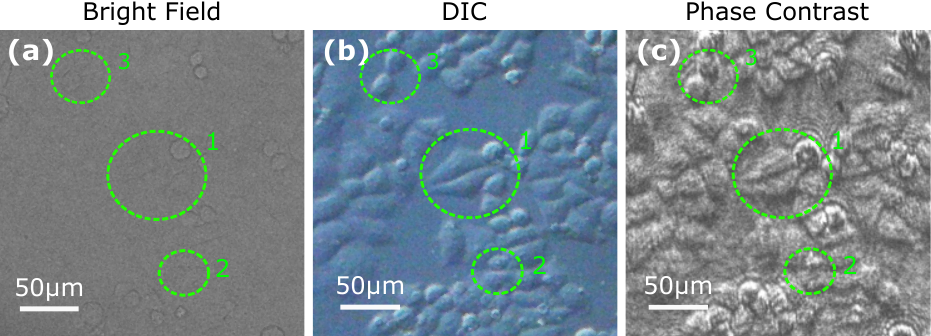}
    \caption{\label{fig:HeLa_DIC}A differential interference contrast microscopy image of the HeLa cells is given in (b). For comparison purposes, its bright field image (a) and experimental phase contrast image obtained with the notch filter (c) are reproduced from the main article.}
\end{figure*}

Bright field and phase contrast microscopy of the HeLa cells were performed using the configuration given in Fig. 5(a) of the main article. First, bright field images were obtained using white light from an inverted microscope (Nikon Ti-80i) without the notch filter. The source was then replaced by fibre-coupled (Thorlabs SM600) laser light (Thorlabs S1FC635 \SI{635}{\nano\meter}) collimated by an objective (Olympus A4 4x/0.1NA), and circularly polarized by a linear polarizer (Thorlabs LPVISC100-MP2) and quarter waveplate (Thorlabs AQWP05M-600). These were contained within an optical cage mounted on an XYZ-stage to vary incident angles. Moreover, the notch filter was place immediately beneath the sample to perform phase imaging. A second microscope objective (Nikon LU Plan 50x/LWD) collected the filtered light onto a camera (Andor Zyla sCMOS 4.2P). Given in Fig. \ref{fig:HeLa_DIC}(b), a differential interference contrast microscopy image was taken on a second microscope (Olympus BX60) using a separate microscope objective (Olympus PlanN 20x/0.4NA).

\subsection{Biological Sample Preparation}

The HeLa cells were grown in Dulbecco’s modified Eagle’s medium (DMEM) (Lonza) supplemented with 10\% heat inactivated bovine growth serum (Gibco), 1x Pen-Strep (Lonza) at \SI{37}{\celsius} with 5\% carbon dioxide. The cells were plated \SI{24}{\hour} before fixation onto \SI{35}{\milli\meter} glass bottom dishes, before being fixed with 4\% paraformaldehyde for \SI{15}{\minute} at room temperature and washed 3 times with phosphate-buffered saline (PBS).



\providecommand{\noopsort}[1]{}\providecommand{\singleletter}[1]{#1}%

\end{document}